\documentclass[aps,prl,superscriptaddress,reprint]{revtex4-1}
\usepackage{amsmath}
\usepackage[pdftex]{graphicx}

\begin{document}
\title{Non-Markovian quantum-classical ratchet for ultrafast long-range electron-hole separation in condensed phases }
\author{Akihito Kato}
\affiliation{Institute for Molecular Science, National Institutes of Natural Sciences, Okazaki 444-8585, Japan}
\author{Akihito Ishizaki}
\affiliation{Institute for Molecular Science, National Institutes of Natural Sciences, Okazaki 444-8585, Japan}
\affiliation{School of Physical Sciences, The Graduate University for Advanced Studies, Okazaki 444-8585, Japan}
\date{\today}

\begin{abstract}
In organic photovoltaic systems, a photogenerated molecular exciton in the donor domain dissociates into a hole and an electron at the donor/acceptor heterojunction, and subsequently separate into free charge carriers that can be extracted as photocurrents. The recombination of the once-separated electron and hole is a major loss mechanism in photovoltaic systems, which controls their performance. Hence, efficient photovoltaic systems need built-in ratchet mechanisms, namely, ultrafast charge separation and retarded charge recombination.
In order to obtain insight into the internal working of the experimentally observed ultrafast long-range charge separation and protection against charge recombination, we theoretically investigate a potential ratchet mechanism arising from the combination of quantum delocalization and its destruction by performing numerically accurate quantum-dynamics calculations on a model system.
It is demonstrated that the non-Markovian effect originating from the slow polaron formation strongly suppresses the electron transfer reaction back to the interfacial charge-transfer state stabilized at the donor/accepter interface and that it plays a critical role in maintaining the long-range electron--hole separation.
\end{abstract}

\maketitle

Great strides in the development of organic solar cells have posed fundamental physical problems in quantum-dynamical phenomena related to the exciton and charge generation and transport in complex molecular systems. Organic photovoltaic (OPV) systems consist of a blend of donor and acceptor organic materials. The photogenerated molecular exciton in the donor domain dissociates at the donor/acceptor heterojunction into a hole and an electron, which subsequently separate into free charge carriers to be extracted as photocurrents. In efficient OPV materials, in which fullerenes are utilized as the acceptor domain, for example, the electron and hole escape from the heterojunction and long-range charge separation (CS) proceeds efficiently \cite{Park_Bulk_2009,ostroverkhova2016organic}.

However, a question naturally arises concerning the physical mechanism of the long-range CS process. The electron and hole are subject to their mutual Coulomb attraction, and hence they may be thought to relax to a bound electron--hole pair localized at the interface, giving rise to an interfacial charge transfer (CT) state. As the dielectric constants of organic materials are typically small, the electron--hole binding energy ($0.20 - 0.50 \,{\rm eV}$
\cite{zhu2009charge,drori2008below,hallermann2010correlation})
can be one order of magnitude larger than the thermal energy at room temperature ($\sim 0.026 \,{\rm eV}$). Consequently, the Coulomb attraction can stabilize the CT state at the interface \cite{deibel2010role}.
For understanding the crucial factors that determine the energy conversion efficiency of organic solar cells, it is important to elucidate the physical mechanisms of how the electron and hole escape from the donor/acceptor interface, surmounting the Coulomb barrier, and produce mobile charge carriers.

Recent spectroscopic measurements revealed that the long-range CS process took place on an ultrafast timescale, within a few hundred femtoseconds
\cite{muntwiler2008coulomb,grancini2013hot,Jailaubekov_Hot_2012,Gelinas:2014gu,Provencher_Direct_2014}.
Over the past few years, two mechanisms have been the subject of controversy. One mechanism proposed that the excess energy of the photogenerated exciton could help the charges surmount the Coulomb barrier, i.e., the so-called hot exciton dissociation mechanism \cite{muntwiler2008coulomb,grancini2013hot,Jailaubekov_Hot_2012}.
The other proposed that the quantum delocalization of the charges could reduce the Coulomb barrier \cite{bakulin2012role,few2014influence}.
Tamura and Burghardt attributed the ultrafast CS process to a hybrid mechanism, namely, the charge delocalization and generation of vibronically hot interfacial CT states
\cite{tamura2013ultrafast}.

However, no broad consensus has been achieved concerning the physical origins of the ultrafast CS process, at present. Another significant question concerns the avoidance of the recombination of the once-separated electron and hole or the relaxation to the interfacial CT state \cite{lakhwani2014bimolecular}.
The recombination is a major loss mechanism in photovoltaic materials, which controls their performance. Hence, efficient photovoltaic systems need
a type of built-in ratchet
mechanism, namely, ultrafast charge separation and retarded charge recombination \cite{beibel2009prl}.
Here, we note that a similar issue was addressed
\cite{Ishizaki_Theoretical_2009,ishizaki2012quantum,Hoyer_Spatial_2012,scholes2017using}
while investigating the underlying mechanisms of the observed long-lived coherence and its potential role in photosynthetic light harvesting
\cite{engel2007evidence}.
The following mechanism was conjectured:
A fast forward transfer involves ballistic evolution, arising from quantum delocalization.
This quantum delocalization facilitates energetic uphill transfer and avoids local energetic traps. However, such free evolution would allow facile backward transfer, away from the destination, and thus, the next process would be decoherence, which localizes the product state.
Then, the backward transfer is suppressed because it requires incoherent hopping that needs to overcome free energy barriers.
These steps are hypothesized to be able to ensure rectifying actions.
Recently, a similar mechanism was discussed for ultrafast long-range charge separation in OPV systems \cite{Gelinas:2014gu,Brdas_Photovoltaic_2016}.

However, the ratchet effect
in the quantum dynamics of exciton and charge transport still remains a speculation, and the relevance has not yet been proven theoretically or experimentally.
In this letter, we theoretically examine a quantum--classical ratchet mechanism that is made possible via the combination of quantum delocalization and its destruction in order to obtain insight into the inner working of ultrafast long-range CS and protection against the charge recombination or relaxation to the CT state stabilized at the donor/acceptor interface.

\begin{figure}
\centering
\includegraphics[width=80mm]{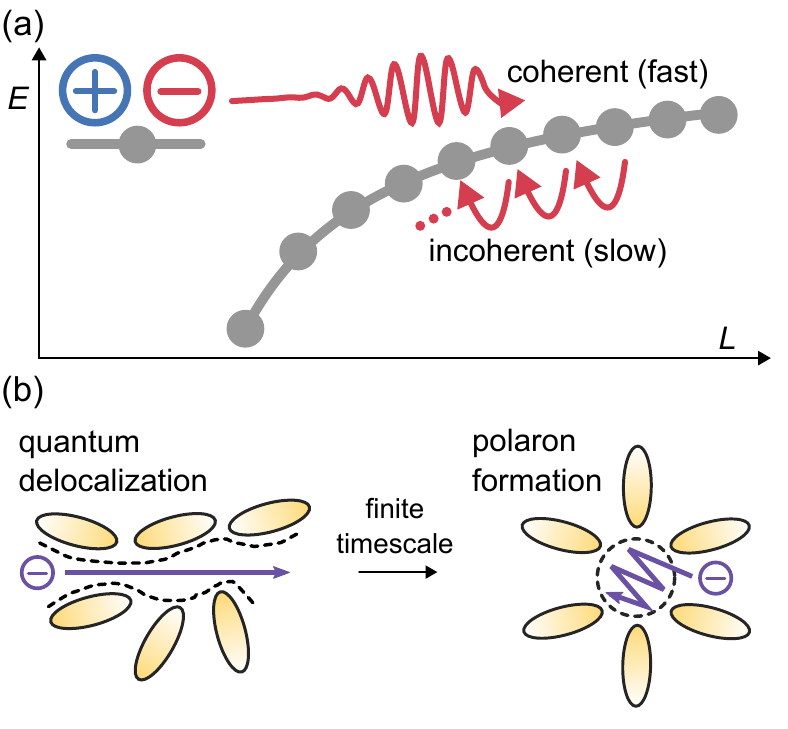}
\caption{Schematics of the charge separation dynamics.
(a) The electron located at the donor/acceptor interface is transferred via quantum delocalization
and long-range charge separation is achieved as a fast process.
The backward propagation of the electron is described by the incoherent hopping motion, which is a slow process. Therefore, the recombination is suppressed by the dynamical coherent--incoherent transition.
(b) The dynamical transition is caused by the small polaron formation, which is induced by the coupling with the surrounding phonon environment.
The timescale of the polaron formation significantly affects the timescale of both the coherent--incoherent transition and the recombination.}
\label{Fig1}
\end{figure}

{\it Model.}---
In OPV systems, mobile charges strongly interact with the organic molecular environment, and thus, the environmental reorganization, or correspondingly, the small polaron formation, proceeds with a finite timescale, where the charges are dressed in environmental phonon clouds. Consequently, the mobility of the charges is reduced \cite{smith2014ultrafast}.
Before the polaron formation is completed, however, quantum delocalization of the mobile charges might be sustained even in the presence of the strong interaction with the environment \cite{Ishizaki_Unified_2009}, and thus, the charges can move freely.
To describe both the dissipationless forward propagation of the charges through quantum delocalization and the small polaron formation with a finite timescale suppressing the relaxation back to the interfacial CT state, it is required to employ a non-Markovian quantum-dynamic theory on the basis of equations of motion for reduced density operators of open quantum systems.

In order to model the photogenerated exciton, the interfacial CT state, and the subsequent CS dynamics, we fix the hole at the interface and consider a finite one-dimensional lattice \cite{Lee:2015fi} as a model of the acceptor domain, where each site corresponds to an acceptor molecule and the lattice constant $\ell$ presents the distance between the contiguous sites.
Site 0 stands for the interface, and the electron at site 0 corresponds to the photogenerated Frenkel exciton.
On the other hand, site $n$ $(n = 1, \dots, N)$ is located at a distance of $L_n=n\ell$ from the interface, and thus, the electron at site $n$ is bounded with the hole via the Coulomb potential, $E_n = -e^2/4\pi\varepsilon_0\varepsilon_{\rm r} L_n$ with $e$, $\varepsilon_0$, and $\varepsilon_{\rm r}$ being the elementary charge, permittivity of vacuum, and dielectric constant, respectively.
See Fig.~\ref{Fig1}.
Throughout this letter, the bounded electron--hole pair states are termed the CT states, for distinguishing them from the free charge carriers.

Let $\lvert n \rangle$ denote the state where the electron is localized at site $n$. Then, the Hamiltonian of the exciton (XT) and CT states is given as
$\hat{H}^{\rm el} = \sum_{n=0}^N E_n \lvert n \rangle \langle n \rvert + \sum_{n=1}^N V_{n, n-1}( \lvert n \rangle \langle n-1 \rvert +{\rm h.c.} )$,
where $V_{10}=V_{\rm XT-CT}$ represents the exciton--CT coupling and $V_{21} = \dots = V_{N, N-1}=V_{\rm CT}$ the CT transfer integrals.
The site energy of the exciton $E_0$ is set to be nearly resonant with the delocalized CT states, which are the eigenstates in the CT manifold. Therefore, the photogenerated exciton at the interface can immediately transfer to a CT state with long-range electron--hole separation.

The phonon Hamiltonian associated with the site $n$ is given by $\hat{H}^{\rm ph}_n = \sum_\xi \hbar \omega_{n\xi} \hat{b}_{n\xi}^+ \hat{b}_{n\xi}$, where $\hat{b}_{n\xi}^+$ and $\hat{b}_{n\xi}$ denote the creation and annihilation operators of the $\xi$th phonon with frequency $\omega_{n\xi}$.
We assume that the interaction between site $n$ and the phonons are given by
$\hat{H}^{\rm el-ph}_n = \lvert n \rangle \langle n \rvert \sum_\xi g_{n\xi}( \hat{b}_{n\xi}^+ + \hat{b}_{n\xi} )$,
where $g_{n\xi}$ stands for the coupling strength of the $\xi$th phonon.
This form of the interaction Hamiltonian induces environmental reorganization or small polaron formation \cite{Jang:2008ef,Ishizaki_Unified_2009}.
The polaron formation and its timescale are characterized by the environmental relaxation function, $\Psi_n(t) = (2/\pi)\int^\infty_0d\omega\,[J_n(\omega) /\omega] \cos\omega t$, where $J_n(\omega)$ stands for the phonon spectral density $J_n(\omega) = \pi \sum_\xi g_{n\xi}^2 \delta(\omega - \omega_{n\xi})$.
Specifically, when the spectral density is given by the Drude--Lorentz form, $J_n(\omega) = 2\lambda_n \tau_n \omega/(\tau_n^2\omega^2+1)$, the relaxation function is expressed as
$\Psi_n(t) = 2\lambda_n \exp(-t/\tau_n)$
with $\lambda_n$ and $\tau_n$ being the environmental reorganization energy and timescale of the polaron formation associated with site $n$, respectively.
When the value of $\tau_n$ is large, the quantum dynamics of the charges exhibit highly non-Markovian behaviors owing to the lack of timescale separation between the dynamics of the charges and the surrounding environment.
To focus on the roles of the reorganization energy $\lambda_n$ and the timescale $\tau_n$, we employ the Drude--Lorenz model for the spectral density.
For simplicity, $\lambda_0 = \dots = \lambda_N \equiv \lambda $ and $\tau_0 = \dots = \tau_N \equiv \tau$ are assumed throughout the letter.

An adequate description of the CS dynamics is provided with the reduced density operator $\hat\rho(t)$, i.e., the partial trace of the density operator of the total system over the phonon degrees of freedom.
The time evolution of the reduced density matrix $\rho_{m n}(t) = \langle m \vert \hat\rho(t) \vert n \rangle$ for the above Hamiltonians and spectral density can be solved in a numerically accurate fashion \cite{Ishizaki_Unified_2009} through the use of the so-called hierarchical equations of motion approach \cite{Tanimura_Stochastic_2006}.
This approach adequately describes the small polaron formation process with finite timescales, which has not been well captured in the approach using the multiconfiguration time-dependent Hartree method \cite{tamura2013ultrafast,Tamura:2012ej,polkehn2018quantum}.
For numerical calculations, the Coulomb binding energy is fixed so that $e^2/4\pi\varepsilon_0\varepsilon_{\rm r} \ell = 0.30\,{\rm eV}$.
The other parameters are set to $N = 10, E_0 = 0, V_{\rm XT-CT}=0.15\,{\rm eV}, V_{\rm CT}=0.10\,{\rm eV}$, $\lambda=0.02\,{\rm eV}$, and $T=300\,{\rm K}$, which are typical in OPV systems \cite{Tamura:2012ej,tamura2013ultrafast,Lee:2015fi,DAvino:2016gp,barker2014distance}.
We checked the consistency of the numerical results, with respect to the role of the non-Markovian polaron formation, by increasing the number of CT sites up to $N=20$.

\begin{figure}
\centering
\includegraphics[width=80mm]{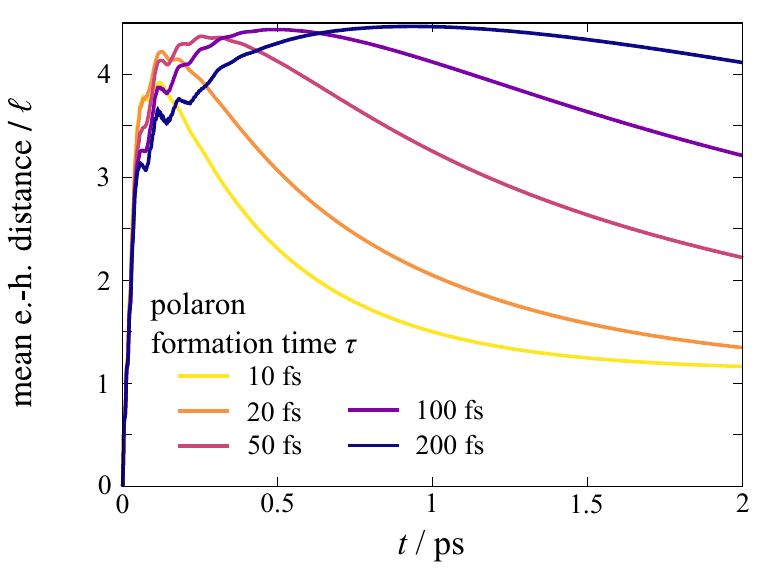}
\caption{Time evolution of the mean electron--hole distance $\langle L_n \rangle_t = \sum_{n=1}^N n\ell\, \rho_{nn}(t)$ for various values of the small polaron formation time $\tau$.}
\label{Fig2}
\end{figure}

{\it Results.}---
To investigate the impacts of the small polaron formation process, on the CS dynamics, we address the mean electron--hole distance.
Figure~\ref{Fig2} presents the time evolution of the mean
electron--hole distance, $\langle L_n \rangle_t = \sum_{n=1}^N n\ell \, \rho_{nn}(t)$, for various values of polaron formation time $\tau$.
In the short-time region of $t<50\,{\rm fs}$, the mean distance increases up to $\langle L_n \rangle_t/\ell \simeq 4$, independent of the polaron formation time. This indicates that the electron travels through the use of quantum delocalization, straddling multiple sites, leading to fast and long-range forward electron transfer. The quantum delocalization facilitates the energetic uphill transfer.

As time increases, however, the mean electron--hole distance $\langle L_n \rangle_t$ exhibits a decrease, indicating a backward and energetically downhill electron transfer to the interfacial CT state.
This turnover is caused by the inherent coherent-to-incoherent transition.
That is, owing to the electron--phonon interaction, the formation of the small polaron proceeds, and the electron falls off into one of the minima of the free energy and becomes localized \cite{Ishizaki_Unified_2009,Ishizaki:2010fx}.
In this situation, the electron transfer is described as incoherent hopping; it generally follows the classical rate law, where the rate depends on the environmental parameters and satisfies the detailed balance condition.
Consequently, the backward electron transfer exhibits the $\tau$-dependence.
Furthermore, Fig.~\ref{Fig2} demonstrates that the slower polaron formation time is capable of suppressing the backward electron transfer more strongly, and thus, can preserve the long-range charge-separated state for a longer time.
This ``non-Markovian protection'' against the transfer back to the interfacial CT state is explained by the so-called dynamical solvent effect in condensed-phase electron-transfer reactions \cite{zusman1980outer,calef1983classical,hynes1986outer}.
In general, the incoherent electron-transfer reactions in condensed phases involve thermal activation for overcoming the reaction barrier or the free energy of activation, and hence, slower environmental fluctuations yield slower thermal activation.
Indeed, the calculated rate of the incoherent backward transfer exhibits $\tau$-dependence of $\propto \tau^{-1}$.
It can be concluded that the present model with the slow polaron formation times captures the experimentally observed ratchet-like behavior of the ultrafast long-range charge separation and suppression of the charge recombination.

\begin{figure}
\centering
\includegraphics[width=80mm]{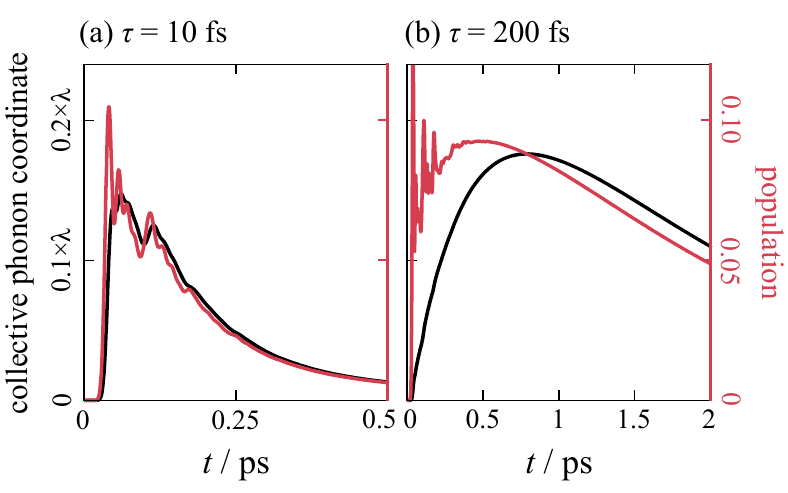}
\caption{Time evolution of the statistically averaged collective phonon coordinate associated with the terminal CT site, $\langle \hat{X}_{N=10}(t) \rangle$, and the time evolution of the electron population at the terminal CT site.
The polaron formation time $\tau$ is set to be (a) $10\,{\rm fs}$ and (b) $200\,{\rm fs}$. }
\label{Fig3}
\end{figure}

To obtain further insight into the non-Markovian aspects in the CS dynamics, we address the collective phonon coordinate defined as $\hat{X}_n \equiv \sum_\xi g_{n\xi}( \hat{b}_{n\xi}^+ + \hat{b}_{n\xi})$, which is also termed the solvation coordinate, in the literature on condensed-phase electron-transfer reactions \cite{sparpaglione1988dielectric}.
Specifically, we consider the time evolution of the statistical average, $\langle \hat{X}_n(t) \rangle \equiv {\rm Tr}[ \hat{X}_n \hat{\rho}^{\rm tot}(t) ]$, where $\hat{\rho}^{\rm tot}(t)$ is the density operator of the entire system.
For the phonon Hamiltonian, $\langle \hat{X}_n(t) \rangle$ is expressed by the relaxation function $\Psi_n(t)$ as $\langle \hat{X}_n(t) \rangle = -\int_0^t ds \, [\partial_s \Psi_n(s)] \rho_{nn}(t-s)$ \cite{Kato:2016bg}.
The value of $\langle \hat{X}_n(t) \rangle$ quantifies the degree of solvation of the electron at site $n$, namely, the degree of the small polaron formation.
In the absence of the electron $\rho_{nn}(t) = 0$, when the site is in an electrically neutral state, each phonon mode is oriented randomly. Such a situation is represented by $\langle \hat{X}_n(t) \rangle = 0$.
When the electron is located at site $n$, $\rho_{nn}(t) = 1$, and all the phonon modes are fully polarized and the small polaron formation is completed ($t\to \infty$). Then, we obtain $\langle \hat{X}_n(t\to\infty) \rangle = -\int_0^\infty ds \, [\partial_s \Psi_n(s)] = \Psi_n(0) = 2\lambda$.
Furthermore, in the Markov limit, the statistically averaged coordinate is approximately expressed as $\langle \hat{X}_n(t) \rangle \simeq - \int_0^\infty ds \, [\partial_s \Psi_n(s)]\cdot \rho_{nn}(t) = 2 \lambda \rho_{nn}(t)$, which corresponds to the instantaneous polaron formation.
Therefore, the non-Markovian nature in the suppression of the backward electron transfer can be ensured by the time-lagged behavior of $\langle \hat{X}_n(t) \rangle$ behind the population $\rho_{nn}(t)$.
Figure~\ref{Fig3} presents the time evolution of the statistically averaged collective phonon coordinate associated with the terminal CT site, $\langle \hat{X}_{N=10}(t) \rangle$.
In the case of fast polaron formation ($\tau=10\,{\rm fs}$) in Fig.~\ref{Fig3}a, the time evolution of $\langle \hat{X}_N(t) \rangle$ follows the population on the terminal CT site, $\rho_{NN}(t)$ instantaneously.
However, Fig.~\ref{Fig3}b for the slow polaron formation case ($\tau=200\,{\rm fs}$) exhibits a lag of the coordinate behind the population.
This time-lagged behavior, which is attributed to the non-Markovian nature, indicates the existence of sequential processes enabling the ratchet mechanism, as speculated above.
That is, (1) the fast forward transfer to the terminal CT site takes place through the use of quantum delocalization. Sequentially, (2) the decoherence to localize the product state proceeds for deterring the facile backward transfer away from the terminal CT site. Thus, (3) the backward transfer is suppressed because of the incoherent hopping dominated by the slow thermal activation, for overcoming the reaction barrier.

{\it Concluding remarks.}---
In this letter, we theoretically investigated the ratchet mechanism that was made possible via the combination of quantum delocalization and its destruction, in order to understand the underlying physical mechanisms of the ultrafast long-range charge separation and protection against charge recombination.
We specifically explored the interplay between the fast forward transfer of the electron through the use of quantum delocalization and the slow backward transfer induced by the small polaron formation, by performing numerically accurate quantum-dynamic calculations.
The non-Markovian effect that originated from the slow timescales of the polaron formation strongly suppressed the backward electron transfer, and consequently, played a critical role in maintaining the large electron--hole distance.
The ratchet mechanism proposed in this letter is not restricted in organic photovoltaic  systems.
This is because the recent progresses in nanotechnology have begun to explore the engineering of the electron--phonon coupling \cite{benyamini2014real}, and pinning down the system in the excess energy states, by protecting from de-excitation, may be experimentally realizable.
The quantum Zeno effect induced by the repeated measurements can also be utilized to control the systems \cite{kondo2016using,hacohen2018incoherent}.
Combinations of the non-Markovian polaron formation with optimized measurement protocols may give us a novel method for controlling and realizing the desired state of quantum systems.

In this study, we neglected the various contributions examined in the previous studies, such as the high-frequency vibrational modes, static disorder, hole delocalization effect, and dimensionality of the system.
Whether inclusion of such contributions will change our understanding of the CS dynamics will be an interesting topic for future discussion. The future works can be focused on the interplay between the quantum coherence and its decoherence effect for the optimization of the charge and energy transport.

\acknowledgments
This work was supported by the JSPS KAKENHI Grant Number 17H02946 and JSPS KAKENHI Grant Number 17H06437 in Innovative Areas ``Innovations for Light-Energy Conversion (I$^4$LEC).''


\end{document}